\documentclass[amssymb,prl,aps,twocolumn,amsmath,showpacs]{revtex4} 
\usepackage[T1]{fontenc}
\usepackage{graphicx}
\usepackage{color}
\usepackage{amsmath}
\usepackage{ulem}

\newcommand{\VR}{V_{\rm{R}}}
\newcommand{\Vs}{V_{\rm{R}}^{*}}

\newcommand{\sigR}{\sigma_{\rm{R}}}

\newcommand{\ls}{l_{\rm{s}}}

\newcommand{\Isat}{I_{\rm{sat}}}
\newcommand{\kL}{k_{\rm{L}}}

\newcommand{\lsi}{l_{\sigma}}
\newcommand{\Esi}{E_{\sigma}}
\newcommand{\tsi}{t_{\sigma}}

\newcommand{\IW}{I_{\rm{W}}}
\newcommand{\tW}{t_{\rm{W}}}

\newcommand{\kbf}{\bf{k}}
\newcommand{\kzerobf}{\bf{k_0}}

\newcommand{\ybf}{\bf{y}}

\newcommand{\sone}{\sigma_{\rm{1}}}
\newcommand{\stwo}{\sigma_{\rm{2}}}

\newcommand{\gradp}{\nabla^2_{\perp}}

\begin{document}

\title{From weak to strong localization : observation of coherent back-scattering and its dynamics in a transverse 2D photonic disorder}

\date{\today}
\author{Julien Armijo$^{1, 2}$}

\email[Corresponding author: ]{julienarmijo@gmail.com}

\author{Rapha\"el Allio$^{1}$}

\affiliation{$^1$Departamento de F\'isica, MSI-Nucleus on Advanced Optics, and Center for Optics and Photonics, Facultad de Ciencias, Universidad de Chile, Santiago, Chile\\
$^2$Facultad de F\'isica, Pontificia Universidad Cat\'olica de Chile, Santiago, Chile}

\begin{abstract}

We report the first observation of coherent back-scattering (CBS) of light in a transverse photonic disorder. The CBS peak is recorded in the far-field, at a fixed propagation time set by our crystal length, and displays a contrast approaching the ideal value of 1, which proves good coherence of transport in our system. We study its dynamics for increasing disorder strength, and find a non-monotonous evolution. For weak disorder, the CBS signal increases, and the asymmetry of the momentum distribution becomes inverted compared to the initial condition. For stronger disorder, we observe a resymmetrization of the momentum distribution, confirmed by numerical simulations, and compatible with the onset of strong (Anderson) localization. 
\end{abstract}

\maketitle


\paragraph{Introduction.}

\begin{figure}
\includegraphics[width=8cm]{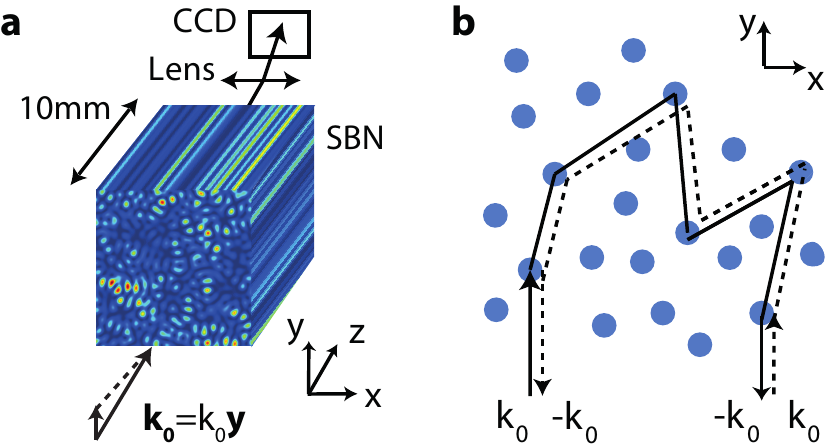}
\caption{(color online) a : Sketch of the experiment. A quasi-plane wave with transverse momentum $k_0$$\ybf$ is launched in a $z$-invariant disorder and recorded at crystal output in Fourier space on a CCD camera.
b : Principle of CBS : all pairs of scattering paths $k_0 \rightarrow -k_0$ (solid line) interfere constructively with their time-reversed counterpart (dashed line).}
\label{fig.sketch}
\end{figure}

Coherent Back Scattering (CBS) is a fundamental effect in mesoscopic physics, by which a wave of initial momentum $\kzerobf$, multiply scattered elastically in a disordered medium, has a probability enhanced by a factor 2 to end in state $-\kzerobf$, compared to any other $\kbf$ within the elastic scattering circle $\vert \kbf \vert = \vert \kzerobf \vert$. This effect is due to constructive interference of the reversed pairs of scattering paths, as shown in Fig. \ref{fig.sketch}.b, and is intimately connected with weak localization (WL) \cite{akkermans07}. It was first predicted for electrons \cite{abrahams79}, and several studies could relate observed phenomena in electron transport, such as negative magnetoresistance, to WL \cite{bergman84}. However, direct observations of the CBS peak were obtained only using macroscopic classical waves such as light \cite{wolf85, vanalbada85}, acoustic waves \cite{bayer93, tourin97}, seismic waves \cite{larose04}, and recently ultracold atoms \cite{jend12}.
Besides proving the coherence of transport, the dynamics of CBS can be used to extract microscopic parameters of the wave propagation in the disorder, in particular the scattering time and phase function \cite{cherroret12, jend12, plisson13}.

The connection between CBS/WL, and strong, i.e., Anderson localization (AL) \cite{anderson58}, which is exponential in real space, has, to our knowledge, remained experimentally unexplored, even though AL has also been observed in many systems \cite{lagendijk09, aspect09}.
Theoretically, it was recently discovered that in Fourier space, besides the CBS peak at $-\kzerobf$, a twin peak at $\kzerobf$, called coherent forward scattering (CFS) peak \cite{karpiuk12, ghosh14, lee14, micklitz14} should emerge at long times due to time reversal symmetry. For systems not spatially bounded, the symmetrization of the momentum distribution with the CFS peak appearance has been proposed as a signature of strong AL \cite{ghosh14, micklitz14}.

In this Letter, we explore the crossover between weak and strong localization, using a 2D photonic disorder generated in a photorefractive crystal, a system where AL has already been studied \cite{schwartz07, boguslawski13}.
We report the first observation of transverse CBS in a photonic disorder, at finite propagation time set by the crystal length, with a peak contrast approaching the ideal value of 1. Further, varying in controlled manner the disorder strength $\VR$ and the incoming plane wave momentum $k_0$, we study the full momentum distributions and the dynamics of the CBS peak, finding a non-monotonous evolution.
In a first stage, the CBS peak grows, while the incoming wave peak decreases, leading to an asymmetric momentum distribution.
This asymmetry reaches a maximum and for stronger disorder, it decays, as expected in the AL regime when the CFS peak appears \cite{karpiuk12, ghosh14}.
Our data do not permit a definite observation of a CFS signal, but we show that the resymmetrization of the momentum distribution is compatible with the onset of AL in our system. Comparison to numerical calculations, showing similar features, supports this scenario.

\begin{figure}[htbp]
\includegraphics[width=8.5cm]{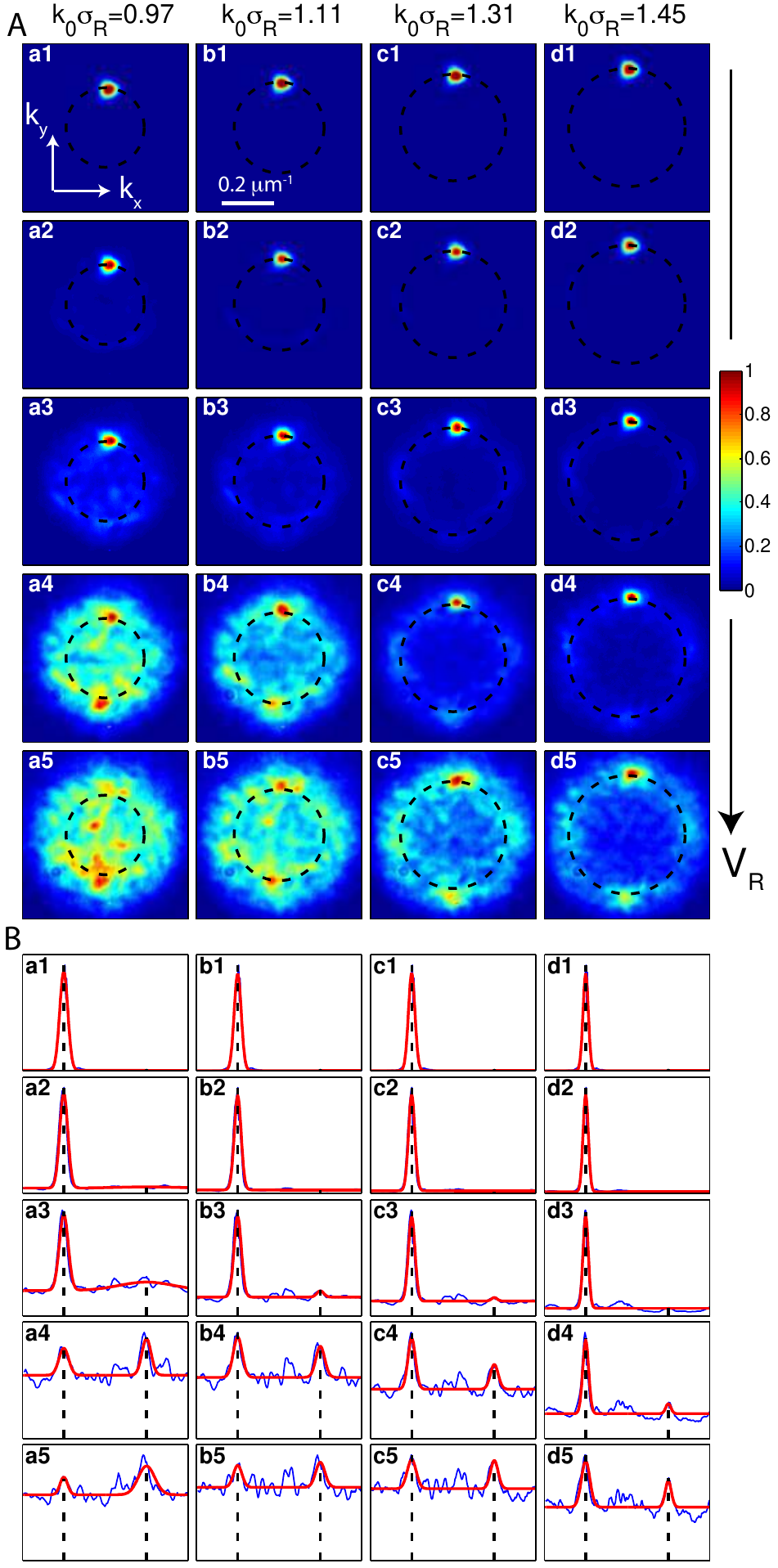}
\caption{(color online) Observation of CBS peak formation (data A) for transverse momenta as indicated on columns, and disorder strengths (from first to last lines) $\VR/\Esi =  0, 0.83, 1.53, 2.12, 2.62$. 
A : full pictures in far-field. Dashed circles show the elastic scattering circles. 
B : Azimuthal profiles. Dashed vertical lines show the positions of the incoming wave at $\theta_1=\pi/2$ and the CBS peak at $\theta_2= 3\pi/2$.}
\label{fig.cbs}
\end{figure}

\paragraph{Our experiments.}
In our setup \cite{armijo14, allio15}, we realize a computer controlled, non-diffracting disorder using spatial light modulators (SLM), as explained in \cite{boguslawski13}. The spectrum of the writing beam is contained in a thin ring (filtered with an SLM in Fourier space), ensuring that the disorder speckle field is almost $z$-invariant through our 10-mm long SBN:75 crystal (see Fig. \ref{fig.sketch}.a), so that the photo-generated disordered refractive index $\delta n(x,y)$ is also $z$-invariant.
The writing beam, at $\lambda=2\pi/\kL = 532$nm, is applied during variable times $\tW$, which allows to adjust the disorder strength $\VR$ \cite{allio15, boguslawski13}, defined as the variance of $\delta n (x, y)$. To calibrate $\VR$, we use our method \cite{armijo14}
\footnote{We assume that the r.m.s refractive index change in a disordered landscape is the same as for a regular lattice as studied in \cite{armijo14}}.
During writing, we apply a bias field is $E_0=+2$kV/cm, so that the disorder is of focusing type (analogous to a red-detuned speckle potential for atoms).
Due to the photorefactive effect \cite{terhalle07, armijo14, allio15}, our disorder is anisotropic, with stronger scattering and localization effects in the c-axis direction $y$.

After writing, we send through the crystal a low intensity, SLM controlled, quasi-plane wave probe beam of adjustable transverse momentum $\kbf=$$k_0$$\ybf$
\footnote{In Data A, the real space r.m.s. width of the probe beam is $\Delta x_{\rm{A}}= 30\mu$m and the measured r.m.s spectral width (see Fig. \ref{fig.cbs}.B) is $\Delta k_{\rm{A}}=0.022/\mu$m, close to the diffraction limit $\Delta k = 1/2\Delta x=0.017/\mu$m. For Data B, the beam width is $\Delta x_{\rm{B}} =120\mu$m but the measured spectral width of $\Delta k_{\rm{B}}=0.015/\mu$m is far from the diffraction limit, probably due to aberrations and imaging resolution.}.
Its slowly varying amplitude $\Psi(x,y)$, in the paraxial regime, obeys a (2+1)D Schr\"odinger equation \cite{deraedt89}
\begin{equation}
i \frac{\partial \Psi}{\partial z} = -\frac{1}{2 \beta_0} \gradp \Psi - \frac{\beta_0}{n_0} \delta n (x, y) \Psi ,
\label{eq.se}
\end{equation}
where $\beta_0=2\pi n_0/\lambda$, $\gradp = \left(\frac{\partial ^2}{\partial x ^2} + \frac{\partial ^2}{\partial y ^2} \right)$ is the transverse laplacian, $z \leftrightarrow t$ plays the role of time $t$, and the potential $V(x,y)$ is the refractive index : $V(x,y) \leftrightarrow - \delta n(x,y)$
\footnote{The correspondence to the Schr\"odinger equation is complete with the additional replacement of the particle mass by the refractive index $m \leftrightarrow n_0$ and the reduced Planck constant $h/2 \pi \leftrightarrow \lambda/2\pi =1/\kL$.}.

All data are averaged over $\sim40$ disorder realizations.
In this work we use a disorder with correlation length $\sigR = 6.53 \mu$m (same definition as for standard speckle disorders with a disk pupil function \cite{kuhn05, piraud12}), which defines a characteristic "time" $\lsi = n_e \kL \sigR^2=  1.17$mm (along $z$), and an "energy" $\Esi = (\kL^2 n_e \sigR^2)^{-1} = 7.2\times 10^{-5}$  \cite{kuhn05}, which we use to normalize the disorder strength $\VR$.

\paragraph{Observation of transverse CBS.}
Figure \ref{fig.cbs}.A shows far-field images of the probe beam after propagation in the transverse disorder, for various $\VR$ and momenta $k_0$. 
For increasing $\VR$, the initial peak at $k_0$ decreases, and one observes the formation, first, of a ring of elastic scattering, then, of a very clear CBS peak at $-k_0$, whose contrast relative to the incoherent background in the scattering circle approaches the ideal value of 1 in many pictures (see, e.g., a4, b4, a5-d5, and color bar).

\begin{figure}[htbp]
\includegraphics[width=8.5cm]{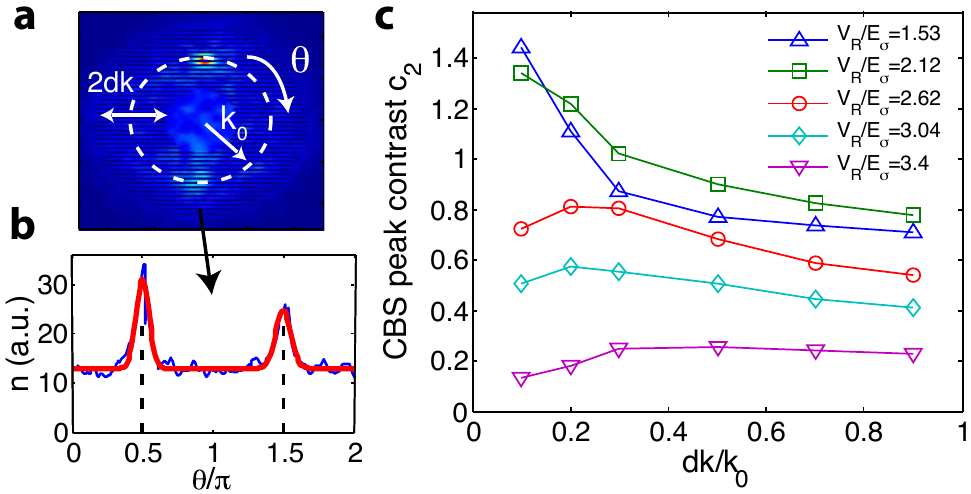}
\caption{(color online) 
Construction of azimuthal profile $n(\theta$), for picture c3 in Data B.
a : Full picture. The dashed area is the integration ring of radius $dk=0.5 k_0$.
b: Resulting $n(\theta)$ with two-peak fit (smooth line). 
c : CBS peak contrast $c_2$ as function of $dk/k_0$, for Data B, c2-c6 ($k_0 \sigR =0.98$).}
\label{fig.ex}
\end{figure}

\begin{figure}
\includegraphics*[width=8.5cm]{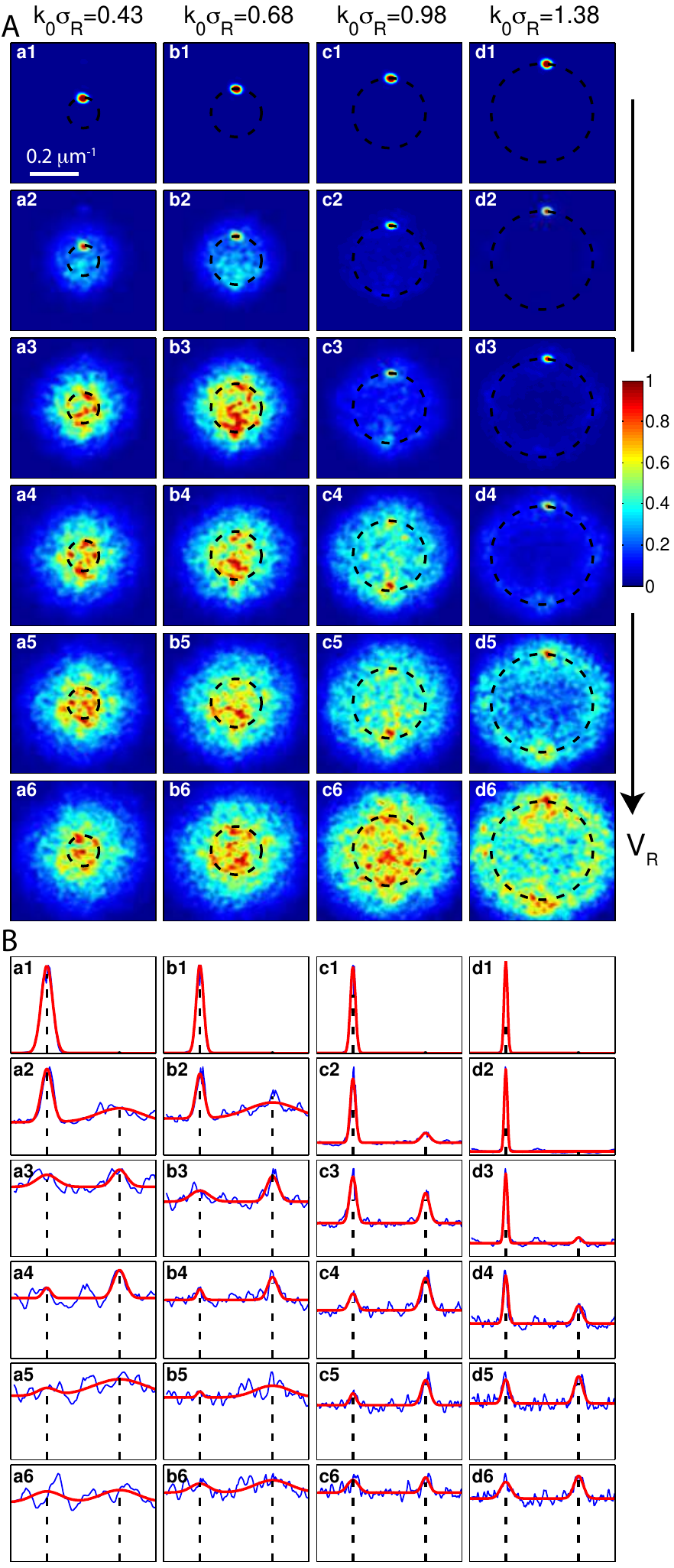}
\caption{(color online) Dynamics of CBS formation (data B) for disorder strengths (from first to last rows) $\VR/\Esi =  0, 1.53, 2.12, 2.62, 3.04, 3.4$. 
}
\label{fig.cbs2}
\end{figure}

For quantitative analysis, we extract azimuthal profiles $n(\theta)$, integrating the density on a ring of radius $dk=0.5 k_0$ (see Fig. \ref{fig.ex}.a). Profiles $n(\theta)$ are then fitted with a two-peak function 
\footnote{Before fitting with Eq. \ref{eq.twopeak}, we allow for an offset $\theta_0$ by fitting the first ($\VR=0$) profile with a one gaussian peak function, then this constant shift $\theta_0$ is applied to all profiles for a given $k_0$.}
\begin{equation}
n(\theta) = n_0 \Big(1 + c_1 e^{- \frac{(\theta-\theta_1)^2}{ 2 \sone^2}} + c_2 e^{- \frac{(\theta-\theta_2)^2 }{2 \stwo^2}} \Big),
\label{eq.twopeak}
\end{equation}
where $\theta_1 = \pi/2$ and $\theta_2 = 3 \pi/2$. In Fig. \ref{fig.ex}.b (data c3 of Data B), the contrast of the CBS peak is $c_2=0.9$. 
It is important however to note that in our system, spectral broadening is strong, so that the scattering circle soon resembles a disk (see Fig. \ref{fig.cbs} and \ref{fig.cbs2}). 
Thus, the choice of $dk$, which is arbitrary, influences the numerical outcome for $n(\theta)$, and the fit.
For example, Fig. \ref{fig.ex}.c illustrates the dependence of $c_2$ on $dk$ (and on $\VR$).
Nevertheless, we have checked that the choice of $dk$ does not qualitatively change our further observations and conclusions.
It can also be noted that the background $n_0$ of incoherent scattering is not isotropic, as in Eq. \ref{eq.twopeak}.
In particular, for the larger $k_0$ one notes clear lobes of preferential scattering directions (see Fig. \ref{fig.cbs}.B.c3-d5), related to the particular autocorrelation function of our disorder \cite{plisson13, kuhn07, piraud13, armijo15}.

\paragraph{Dynamics of momentum distribution asymmetry.}

\begin{figure*}
\includegraphics[width=18cm]{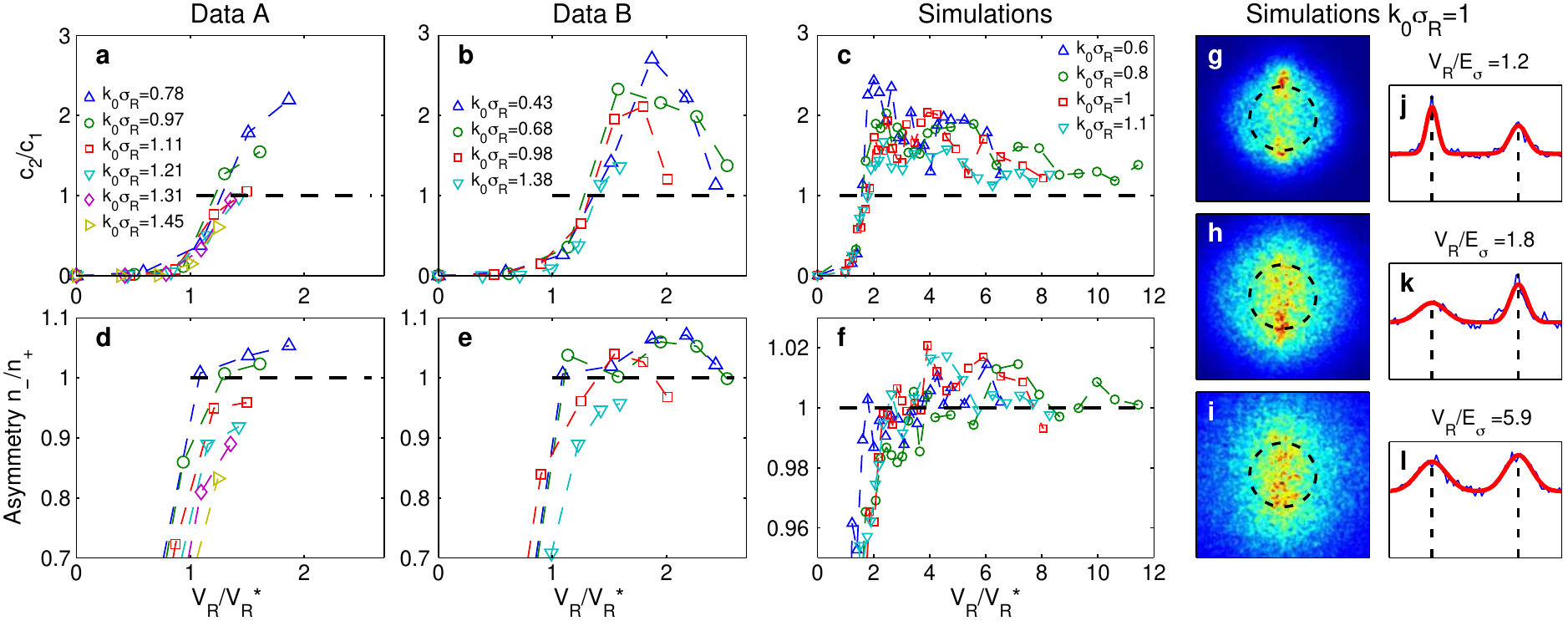}
\caption{(color online) Dynamics of the asymmetry of contrasts $c_2/c_1$ (a-c) and of the momentum distribution $n_+/n_-$ (d-f) in data sets A, B, and simulations, for disorder strengths $\VR$ normalized to $\Vs$ (see text).
Data A : $\Vs/\Esi = 1.41, 1.63, 1.76, 1.84, 1.94, 2.12, 2.45$.
Data B : $\Vs/\Esi  = 1.4, 1.35, 1.7, 2.15$.
Simulations : $\Vs/\Esi =  0.51, 0.61, 0.74, 0.85$.
(g-l) : Examples of simulations results for $k_0 \sigR =1$.
}
\label{fig.asym}
\end{figure*}

To explore the crossover form weak to strong localization, we now focus on the asymmetry of the momentum distribution.
Several works have reported momentum distributions where the asymmetry is opposite to the initial situation, due to the presence of a CBS peak  at $-k_0$ on top of an isotropic scattering ring (e.g., \cite{jend12, labeyrie12}, and for simulations \cite{cherroret12, plisson13}).
However, for strong localization (AL), one expects a resymmetrization of the momentum distribution, which should converge towards a twin peak structure, with CBS and CFS peaks of identical contrast and width \cite{ghosh14}.

In Fig. \ref{fig.cbs2} we show a second data set (B), where the probe beam has a 16 times wider area [33], and $\VR$ reaches higher values.
In a first stage, we observe the formation of well asymmetric profiles with a CBS peak dominating over the initial $k_0$ peak, and $c_2 >c_1$ (see, e.g., a4, a5, b3, b4, c4, c5). For larger $\VR$, the CBS peak decreases and the distribution recovers symmetry, with $c_1 \simeq c_2$.
(see, e.g., a6, b6, c6).

To quantify these observations, Figure \ref{fig.asym} gathers the evolutions of two indicators of asymmetry : the ratio of contrasts $c_2/c_1$, and the ratio $n_-/n_+$ of the total density in the lower ($k_y<0$) \textit{vs} upper half ($k_y>0$) of the momentum space.
The first ratio indicates the relative weight of $-k_0$ \textit{vs} $k_0$ components, independently from a decrease of both $c_1$ and $c_2$ at strong disorder (as seen in Fig. \ref{fig.ex}.c and Fig. \ref{fig.cbs2}).
The second ratio is more general.
To compare all our data at different $k_0$, we rescale the $\VR$ using the value $\Vs$ that we define as the value for which the amplitude $n_1=n_0 c_1$ of the initial peak has decayed to 10\% of its initial value. 
This rescaling makes all curves for different $k_0$ collapse onto a typical path. 
Physically, this procedure removes the $k$-dependence of the scattering mean free paths $ \ls(k)$, i.e., one essentially has $\Vs(k) \sim \ls(k)^{-1}$.
In data A (Fig. \ref{fig.asym}.a, d), both asymmetry ratios increase. In data B (Fig. \ref{fig.asym}.b, e), both reach a maximal value (about 2 for $c_2/c_1$ and 1.07 for $n_+/n_-$), and then decrease towards approximately 1, meaning full resymmetrization at the maximal $\VR$.

\paragraph{Comparison to numerical calculations.}
To test our observations, we simulate Eq. \ref{eq.se}, using a split-step beam propagation algorithm, and a gaussian probe beam of r.m.s width $\Delta x_{\rm{sim}}=200\mu$m. For the disordered $\delta n(x,y)$, we solve the standard steady-state anisotropic equations for the photorefractive effect (see \cite{desyatnikov05} or Eq. 9 in \cite{allio15}).
Yet, an important difference should be noted.
In the experiments, the disorder strength $\VR$ is varied by adjusting the photorefractive writing time, while the writing intensity $\IW$ is constant and much larger than the saturation intensity ($\IW/\Isat \sim 10^3$) \cite{armijo14}. Whereas, in the numerics, steady-state is assumed ($\tW = \infty$) and $\VR$ is adjusted by varying $\IW/\Isat = 0.1 - 5$, i.e., from weak to considerable conditions of photorefractive saturation.

We average the simulations over 50 disorder realizations, and perform the same data analysis as for the experiments.
Overall, the simulations (Fig. \ref{fig.asym}.c,f) display similar trends as the experiments for both asymmetry ratios, with first a maximum (also $\sim 2$ for $c_2/c_1$, but a lower $\sim 1.02$ for $n_+/n_-$), and then a decay, which is much slower than in the experiment.
In Fig. \ref{fig.asym}.g-l, we show examples of the simulation results for $k_0 \sigR=1$. In the first case (g,j), the $k_0$ peak dominates. In the intermediate case (h,k), the CBS peak dominates, and maximal asymmetry is reached. In the third case (i,l), the asymmetry has almost disappeared. There, one also notes a strong disorder broadening, as also seen in the experiment (see Fig. \ref{fig.cbs2}.A.c6), and a strong anisotropy in the momentum distribution, more pronouced than in the experiment.
Despite several quantitative differences, which may well be due to the above-mentioned difference of disorder type and saturation effects, one notes that the maximum asymmetry occurs for similar absolute values of $\VR$, for example, for $c_2/c_1$ and $k_0\simeq 1$, at $\VR \simeq 2.4 \Esi$ in data B and $\VR \simeq 2.6 \Esi$ in simulations. In both cases also, the maximum of $n_+/n_-$ occurs for slightly larger $\VR$. And we neither found any sharp CFS peak in the simulations.
Most importantly, the similarity between measured and simulated dynamics, allows us to   consider the resymmetrization as a fully coherent effect, given that lattice imperfections (for example residual $z$-variations of the disorder), are absent in the numerics.

Can we interpret the resymmetrization at stronger disorder to the onset of strong (Anderson) localization ?
One could firstly note that exponential localization in real space, interpreted as AL, has been well observed in essentially the same system for similar parameters \cite{schwartz07, boguslawski13, armijo15}.
Analysing our parameters more quantitatively brings further insight.
Our measurements are all carried at fixed propagation "time" $t / \tsi \simeq 8.55$.
In the numerical work \cite{karpiuk12}, using a similar disorder (although the pupil function is a disk instead of a ring in our case), the timescale for the CFS peak contrast to reach 0.2 is $t/\tsi \sim 20$ for $\VR=3-4 \Esi$, and $k_0 \sigR = 1.5$. 
Following our rescaling procedure, in our Data B with similar $\VR =3.4 \sigR$, for $k_0 \sigR \simeq 0.8$, one expects similar effects for $t/\tsi \sim 11$
\footnote{To rescale the times we use $\Vs(k) \sim \ls(k)^{-1}$ and in Data A the values $\Vs = 2.45 \Esi$ for $k_0 \sigR = 1.45$ and $\Vs = 1.41 \Esi$ for $k_0 \sigR = 0.78$.}, which is fairly compatible with our parameters. It thus seems very reasonable to attribute the resymmetrization of the momentum distribution at finite time in our data, to the crossover form weak to strong localization.

\paragraph{Conclusion.}
We have reported the first observation of transverse CBS and full 2D momentum distributions in a photonic disorder.
We studied the dynamics of the CBS peak and of the momentum distribution asymmetry.
We noted, first, a growth of the CBS peak, whose contrast approaches 1, and then for stronger disorder, a resymmetrization, which we attributed to the onset of Anderson localization.
Further analysis could provide better understanding of scattering, spectral broadening, and coherence in our disorder, and better estimates of the localization parameters, which may allow to resolve the CFS peak appearance in our system.

\begin{acknowledgments}
We acknowledge helpful discussions with L. Sanchez-Palencia and we thank M. Boguslawski for discussions and for lending us a propagation code and a relaxation algorithm to simulate photorefractive disorders.
Work supported by Programa de Financiamiento Basal de CONICYT (Grant FB0824/2008), Pograma ICM (Grant P10-030-F) and FONDECYT grant 3150587.
\end{acknowledgments}

\bibliographystyle{prsty}

\end{document}